\documentclass[journal]{IEEEtran}
\usepackage{amsmath,amsfonts}
\usepackage{algorithmic}
\usepackage{array}
\usepackage[caption=false,font=normalsize,labelfont=sf,textfont=sf]{subfig}
\usepackage{textcomp}
\usepackage{stfloats}
\usepackage{url}
\usepackage{verbatim}
\usepackage{graphicx}
\usepackage{cite}
\usepackage{makecell}
\usepackage{array}
\usepackage{multirow}
\usepackage{graphbox}
\usepackage{endnotes}
\usepackage{color}
\usepackage[table]{xcolor}
\usepackage{array}
\usepackage{xcolor}

\usepackage[utf8]{inputenc} 
\usepackage[T1]{fontenc}    
\usepackage{hyperref}       
\usepackage{url}            
\usepackage{booktabs}       
\usepackage{amsfonts}       
\usepackage{nicefrac}       

\usepackage{amsmath}
\usepackage{multirow}
\usepackage{bm}
\usepackage{paralist} 
\usepackage{graphicx}
\usepackage{wrapfig}
\usepackage{bbding}
\usepackage{multicol}
\usepackage{color}
\usepackage{amssymb}
\definecolor{mygreen}{HTML}{39b54a}  

\def\BibTeX{{\rm B\kern-.05em{\sc i\kern-.025em b}\kern-.08em
    T\kern-.1667em\lower.7ex\hbox{E}\kern-.125emX}}
\usepackage{balance}
\begin{document}
\newcolumntype{L}[1]{>{\raggedright\arraybackslash}p{#1}}
\newcolumntype{C}[1]{>{\centering\arraybackslash}p{#1}}
\newcolumntype{R}[1]{>{\raggedleft\arraybackslash}p{#1}}

\title{Multi-granularity Backprojection Transformer for Remote Sensing Image Super-Resolution}

  \author{Jinglei Hao,
        Wukai Li, 
        Binglu Wang$\dagger$,
	  Shunzhou Wang,
        Yuting Lu,
        Ning Li, and Yongqiang Zhao,\\
        \thanks{Jinglei Hao, Wukai Li, Yuting Lu and Yongqiang Zhao are with School of Automation, Northwestern Polytechnical University, Xi’an 710072, China (e-mail:haojinglei@mail.nwpu.edu.cn, liwukai51@gmail.com, lyt1996@nwpu.edu.cn, zhaoyq@nwpu.edu.cn).}
      
		\thanks{Binglu Wang is with the Radar Research Laboratory, School of Information and Electronics, Beijing Institute of Technology,Beijing 100081, China  (e-mail: wbl921129@gmail.com).( $\dagger$Corresponding author: Binglu Wang.)}
		\thanks{Shunzhou Wang is with School of Electronic and Computer Engineering, Shenzhen Graduate School, Peking University, Shenzhen, 518055, China. (shunzhouwang@163.com).}
        \thanks{Ning Li is with the Department of Automation, Tsinghua University, Beijing 100084, China (email: lining2022@tsinghua.org.cn). }
  
        \thanks{Our code will be available at https://github.com/KarlLi5/MBT.}

  }


\maketitle

\begin{abstract}
Backprojection networks have achieved promising super-resolution performance for nature images but not well be explored in the remote sensing image super-resolution (RSISR) field due to the high computation costs. In this paper, we propose a Multi-granularity Backprojection Transformer termed MBT for RSISR. MBT incorporates the backprojection learning strategy into a Transformer framework. It consists of Scale-aware Backprojection-based Transformer Layers (SPTLs) for scale-aware low-resolution feature learning and Context-aware Backprojection-based Transformer Blocks (CPTBs) for hierarchical feature learning. A backprojection-based reconstruction module (PRM) is also introduced to enhance the hierarchical features for image reconstruction. MBT stands out by efficiently learning low-resolution features without excessive modules for high-resolution processing, resulting in lower computational resources. Experiment results on UCMerced and AID datasets demonstrate that MBT obtains state-of-the-art results compared to other leading methods.
\end{abstract}


\section{Introduction}
\IEEEPARstart{R}{emote} Sensing Image Super-Resolution (RSISR) is a classical image processing task that aims to reconstruct high-resolution remote sensing images from the low-resolution input. It can be used in many remote sensing interpretation tasks like object detection~\cite{han2014object,li2020object,wang2021multiple} and scene recognition~\cite{cheng2017remote,cheng2018deep}. Thus, many researchers spare no effort to improve the final RSISR performance.

With the renaissance of neural networks, deep learning-based methods have dominated the RSISR field. 
Mainstream RSISR networks rely on feedforward structures for feature extraction, with the primary emphasis on designing more rational learning modules. {For example, Lei et al.~\cite{lei2017super} proposed LGCNet to learn residuals between low-resolution and high-resolution images. Lei et al.~\cite{lei2021hybrid} introduce a novel technique that leverages internal recursion of single-scale and cross-scale information. Wang et al.~\cite{wang2023hybrid} developed a U-shaped structure that aims to further exploit and learn feature relationships among multi-scale remote sensing images.} However, feedforward learning offers only limited contextual information, which encounters challenges in capturing complex textures in images, dealing with high-frequency details, and modeling intricate inter-pixel relationships. 

Different from feedforward learning, DBPN\cite{haris2018deep} ingeniously integrates the backprojection learning into network modules, creating both upsampling and downsampling backprojection blocks. The backprojection blocks incorporate feedback connections, facilitating information propagation within the network. This capability enables the network to delve deeper into the dependencies between low-resolution and high-resolution images. Subsequent work HDPN~\cite{liu2019hierarchical} has further improved the backprojection block, using two $1\times 1$ convolutional layers inside the block to fine-tune LR and HR features. However, the cascading backprojection blocks entail substantial parameter learning for upsampled high-resolution features, leading to a considerable parameter burden. As a result, the exploration of backprojection structures in the field of RSISR has not been well explored in recent years.

To this end, inspired by the success of Transformer for super-resolution~\cite{liang2021swinir,lei2021transformer}, we design a Transformer-based RSISR method termed Multi-granularity Backprojection Transformer (MBT). Specifically, we employ the backprojection learning strategy to learn low-resolution feature representations at different granularities. Firstly, we design the scale-aware backprojection-based Transformer Layer (SPTL), which utilizes pyramid pooling and backprojection mechanisms to learn scale-aware low-resolution features. Based on SPTLs, we construct the context-aware backprojection-based Transformer block (CPTB) for efficient hierarchical feature learning in the network. We organize multiple CPTBs in a cascaded manner to learn the comprehensive low-resolution feature representation of remote sensing images. Moreover, we propose a backprojection-based reconstruction module (PRM) that adopts the backprojection design to learn the differences between high and low-resolution image features, enhancing the hierarchical features for final super-resolution reconstruction. With the above-proposed components, MBT stands out among other backprojection networks as it does not employ excessive modules to process high-resolution features. Therefore, MBT does not consume excessive computational resources. We conduct experiments on commonly-used remote sensing image super-resolution datasets. Our method achieves the best performance in terms of qualitative and quantitative results compared to other state-of-the-art RSISR methods.



To summarize, the contributions of this paper are three-fold:
\begin{itemize}
    \item We develop an SPTL, which can obtain the scale-aware low-resolution features in an effective way.
    \item We propose a CPTB, which can generate comprehensive hierarchical features for complex scene high-resolution image reconstruction.
    \item We construct an MBT, which achieves state-of-the-art results on commonly-used RSISR datasets.
\end{itemize}

The rest of this paper is organized as follows. We will first review the related works in Section II. Then, the details of our proposed method are illustrated in Section III. Experiment results and analysis are introduced in Section IV, and we will give a conclusion of this paper in Section V.

\section{RELATED WORK}
\subsection{Natural Image Super-Resolution}
 Over the past years, through further exploration of end-to-end feature learning, many works in super-resolution reconstruction have further improved reconstruction quality. Most of these works focus on more sophisticated and efficient structural designs. For instance, Zhang et al.\cite{zhang2018residual,zhang2018image} employed densely connected residual blocks to enhance the deep feature learning capacity of networks. Shi et al.\cite{shi2016real} introduced a more efficient upsampling module, allowing the network to perform feature learning on low-resolution images, reducing the parameter burden. This approach is still widely used in various networks. At the same time, generative models have made significant progress in the field of image super-resolution. To achieve better perceptual quality in image reconstruction, several studies have incorporated GAN models to recover more realistic texture details, such as SRGAN\cite{ledig2017photo} and ESRGAN\cite{wang2018esrgan}. Gao et al.\cite{gao2023implicit} introduced diffusion models into the super-resolution domain for continuous image super-resolution reconstruction. Yao et al.\cite{yao2023local} realized high-resolution image generation with a sense of realism at arbitrary scales using flow-based super-resolution models.

\subsection{Optical Remote Sensing Image Super-Resolution}
With the proliferation of neural networks in the realm of super-resolution, numerous endeavors have begun employing neural networks for remote sensing image reconstruction. Lei et al.\cite{lei2017super} proposed LGCNet to learn residuals between low-resolution and high-resolution images. Haut et al.\cite{haut2019remote}, in their residual-based network design, integrated visual attention mechanisms to focus the remote sensing image super-resolution process on deep features demanding greater computational resources. In pursuit of further elevating the quality of remote sensing image reconstruction, Li et al.\cite{li2021single} introduced the SRAGAN network using a GAN model, concurrently applying local and global attention mechanisms to capture features across different scales. In recent years, certain works have gradually shifted focus towards designing more rational structures to explicitly or implicitly capture multi-scale features beneficial for high-resolution reconstruction in remote sensing images. HSENet\cite{lei2021hybrid} effectively exploits internal recursion of single-scale and cross-scale information by employing multi-level feature extraction. Wang et al.\cite{wang2023hybrid} devised a U-shaped structure to further mine and learn feature relationships between multi-scale remote sensing images, enhancing global feature representation through attention based on hybrid convolutions.

\subsection{Residual Back-Projection for Image Super-Resolution}
The feed-forward architecture of deep super-resolution networks achieves promising results in the image super-resolution field. However, the mutual dependency between low-resolution and high-resolution images has not been effectively explored. The iterative back-projection algorithm\cite{irani1991improving} stands as one of the early SR algorithms. It can iteratively calculate the reconstruction error and then fuse it back to adjust the strength of the HR image, aiding in better understanding the image context and enhancing reconstruction quality. In the rapidly evolving field of deep learning, some researchers have focused on exploring how to incorporate traditional back-projection methods into structural designs to harness their full potential for feedback learning in the context of super-resolution. Haris et al.\cite{haris2018deep} introduced back-projection into the SR network architecture, proposing the DBPN network that focuses on employing multiple upsampling and downsampling stages to directly enhance SR features, iteratively learning LR and HR feature maps for feedback residual. Liu et al.\cite{liu2019hierarchical} further improved DBPN, presenting an enhanced back-projection block and implementing bottom-up and top-down feature learning using an HourGlass structure. Additionally, in another work in the same year, Liu et al.\cite{liu2019image} integrated attention into the back-projection module, achieving efficient single-image super-resolution and designing an improved refined back projection block to further enhance SR performance in the final reconstruction process. In recent works, the back-projection block has also been employed to alleviate feature loss caused by the upsampling and downsampling processes. RefSR-VAE\cite{liu2019reference} employs multiple back-projection modules instead of conventional PixelShuffle operations to obtain the final upsampled reconstructed images. 

\subsection{Transformer for Image Super-Resolution}
For low-level visual tasks like super-resolution reconstruction, the inherent computational complexities of the Transformer bring substantial computation burdens. Hence, Liang et al.\cite{liang2021swinir} introduced the Swin Transformer\cite{liu2021swin} into the design of super-resolution networks, proposing an efficient SwinIR that combines the advantages of CNNs and employs a low-burden sliding window for global dependency modeling, achieving impressive results. However, the window attention design for local modeling restricts the powerful global modeling capability of the Transformer. It is evident that increasing the window size can further exploit the Transformer's long-range modeling ability to enhance the quality of reconstructed images, but it also introduces a significant computational burden. To address this issue, Zhang et al.\cite{zhang2022efficient} conducted a more refined design of the SwinIR architecture tailored to low-level visual tasks, proposing ELAN by using windows of different sizes and implementing window attention with larger windows. Chen et al.\cite{chen2023activating} proposed an Overlapping Cross Attention Module to enhance the interaction of cross-window information, effectively activating more pixels for local feature reconstruction. Zhou et al.\cite{zhou2023srformer} introduced Permuted Self-Attention, striking a balance between channel and spatial information in self-attention. By sacrificing channel dimensions in the self-attention computation, this method enables super-resolution networks to enjoy the benefits of large-window self-attention. In addition to addressing the computational burden, it is noteworthy that in SR tasks, Transformers exhibit a tendency to overly emphasize the learning of low-frequency features, which is detrimental to achieving finer reconstruction results. Chen et al.\cite{chen2023activating} addressed this by introducing a CNN branch for the learning of local features and enhancing high-frequency features. Furthermore, Li et al. \cite{li2023feature}. designed a more refined architecture to synergize the learning capabilities of CNN and Transformers, fully leveraging their respective strengths. 

\begin{figure*}[!t]
\centering
\includegraphics[width=\textwidth]{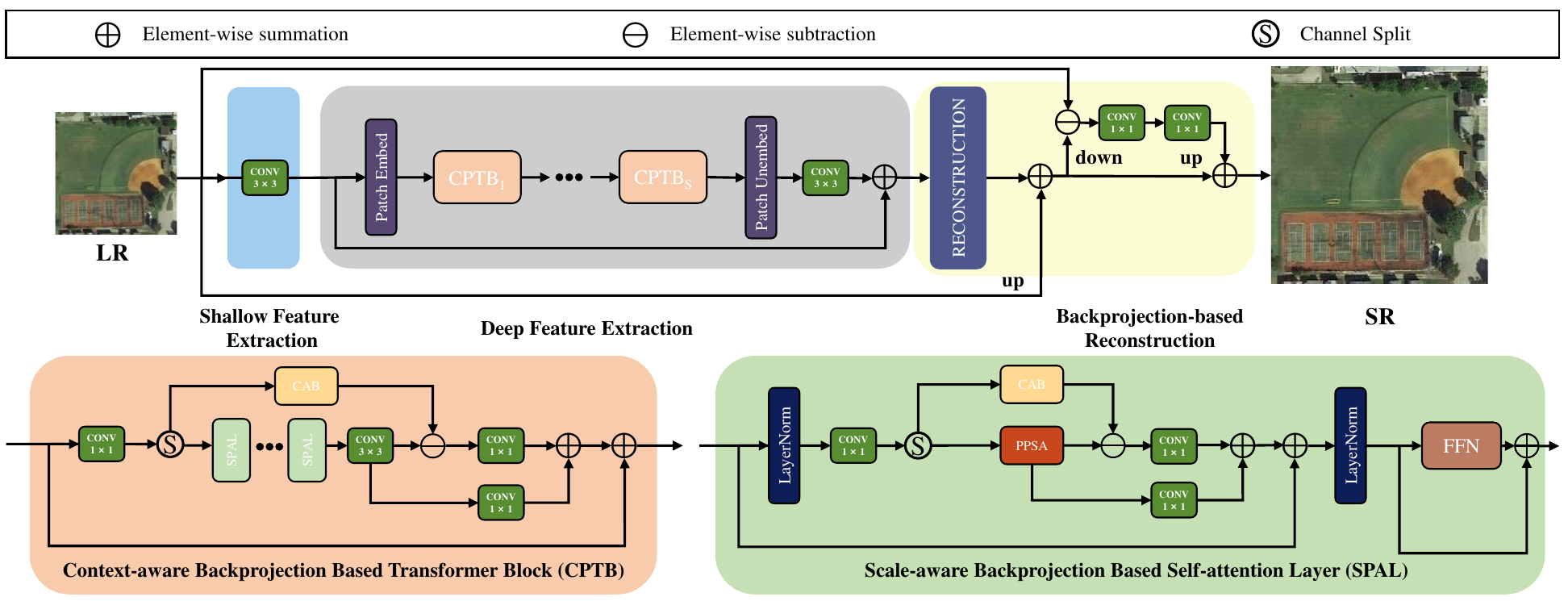}
\caption{The overall architecture of the proposed MBT framework.}
\label{fig_net}
\end{figure*}

\section{METHODOLOGY}
In this section, we introduce the details of MBT. First, we provide a brief overview of the network's overall architecture. Subsequently, in Sections III-B, III-C, and III-D, we delve into the details of the core components of MBT, which include three back-projection structures designed for feature enhancement at different granularities.

\subsection{Overview}
The overall framework of MBT is shown in Fig.\ref{fig_net}. It is composed of three distinct components: shallow feature extraction module, deep feature extraction module, and backprojection-based reconstruction module. Specifically, for the low-resolution input image $\bm{I}_{LR} \in\mathbb{R}^{H\times W \times 3}$, MBT initially employs a $3\times 3$ convolutional layer to extract shallow features $\bm{F}_0 \in\mathbb{R}^{H\times W \times C}$, where $C$ represents the channel dimension. Then, deep feature extraction $\mathcal{F}_{\text{EXTA}}$ is performed through a series of context-aware backprojection-based transformer blocks (CPTBs), followed by obtaining the deep features $\bm{H}^{N} \in\mathbb{R}^{H\times W \times C}$ using a 3$\times$3 convolutional layer of the last CPTB output. Subsequently, we employ a global residual connection to ease training complexity and obtain the initial super-resolved output $\bm{\hat{I}}_{SR}\in\mathbb{R}^{(r\times H)\times (r\times W)\times C}$ through the reconstruction layer $\mathcal{F}_{\text{REC}}$ as 
\begin{equation}
     \bm{\hat{I}}_{SR}=\mathcal{F}_{\text{BINR}}(\bm{I}_{LR})+\mathcal{F}_{\text{REC}}(\bm{H}^{N}).
\end{equation}
Here, $r$ represents the upsampling factor. Lastly, the $\bm{\hat{I}}_{SR}$ is fed into the backprojection-based reconstruction module (PRM) to obtain the final enhanced super-resolved output $\bm{{I}}_{SR}\in\mathbb{R}^{(r\times H)\times (r\times W)\times C}$. 

Among them, CPTB and PRM are the main components of MBT. Moreover, the scale-aware backprojection-based Transformer layer (SPTL) constructs the CPTB. Thus, for the rest of this section, we will first introduce the details of SPTL. Then, CPTB is illustrated. Finally, the details of PRM are presented.


\subsection{Scale-aware backprojection based Transformer Layer}

\begin{figure}[!t]
\centering
\includegraphics[width=0.6\linewidth]{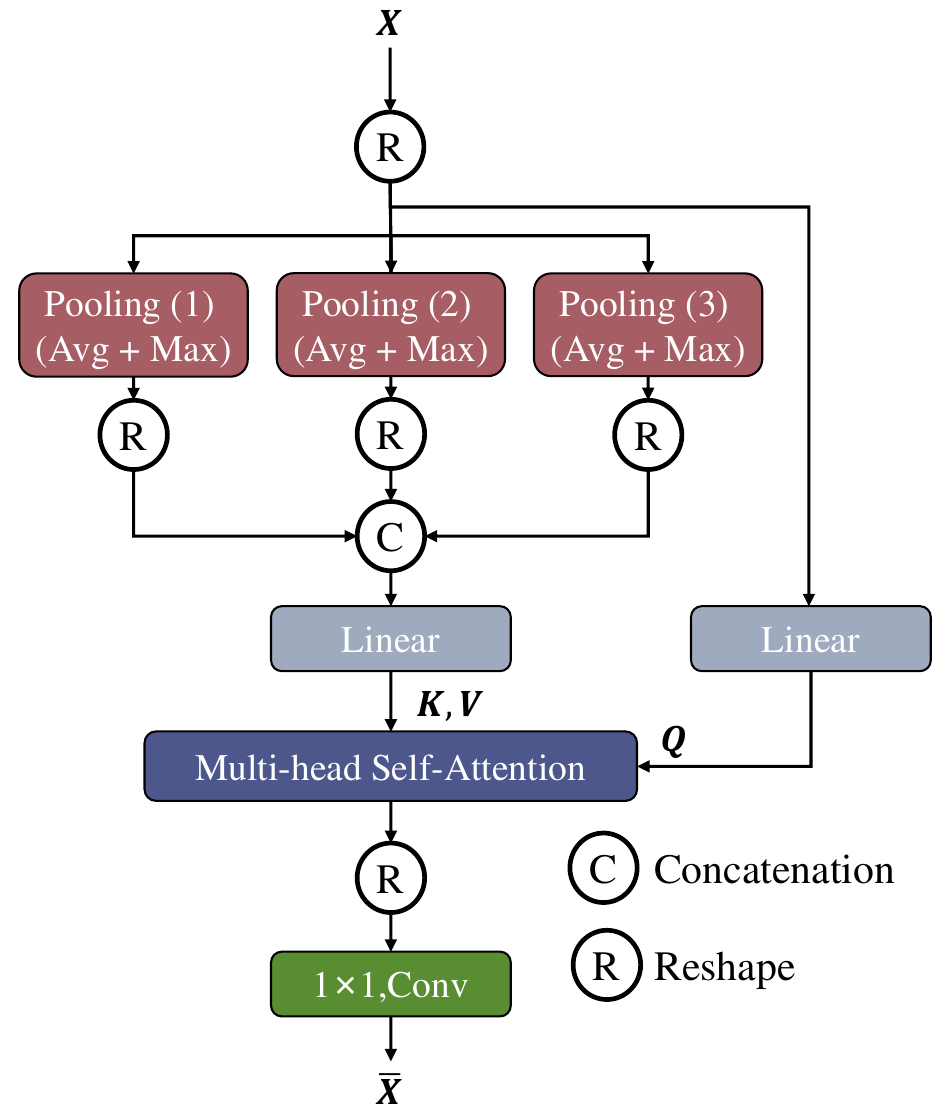}
\caption{Illustration of Pyramid Pooling Self-Attention layer.}
\label{fig_ppsa}
\end{figure}

Extracting multi-scale information in an image can effectively improve the performance of RSISR methods. Therefore, inspired by the design of Pyramid Pooling Transformer~\cite{wu2022p2t}, we propose a multi-scale self-attention operation called Pyramid Pooling Self-attention (PPSA), as shown in Fig.\ref{fig_ppsa}. Specifically, the given input features $\bm{X}\in\mathbb{R}^{H\times W \times C}$ undergo downsampling at different scales through three different scaling pooling blocks\footnote{For the sake of simplicity, we omitted the index number of PPSA.}. Considering that using only average pooling for feature extraction would lead to overly flat reconstruction images with poor contour feature recovery. Thus, we perform a combination of maximum pooling and average pooling operation as the pooling blocks to process the given feature $\bm{X}$ as: 
\begin{equation}
    \bm{P}^{i}=\mathcal{F}_{\text{avg}}^{i}(\bm{X})+\mathcal{F}_{\text{max}}^{i}(\bm{X}),
\end{equation}
where $\bm{P}^{i}$ is the generated specific-scale features and the $i$ denotes the number of pooling blocks.
Here, we opt to use three scales for multi-scale feature extraction, with downsampling ratios of $\times$2, $\times$4, and $\times$8. 

After that, the generated features are concatenated and obtain the multi-scale features $\bm{P}$, and then $\bm{K}$ and $\bm{V}$ values are obtained from $\bm{P}$ through the linear mapping transformation. Subsequently, a standard self-attention calculations $\mathcal{F}_{\text{SA}}$ are performed with the $\bm{Q}$ values obtained from the mapping input features $\bm{X}$, and the output are further processed by a $1\times1$ convolution layers for subsequent feature processing as
\begin{equation}
    \bm{\bar{X}}=\mathcal{F}_{1\times1}(\mathcal{F}_{\text{SA}}(\bm{Q},\bm{K}, \bm{V})).
\end{equation}

\begin{figure}[!t]
\centering
\includegraphics[width=0.5\linewidth]{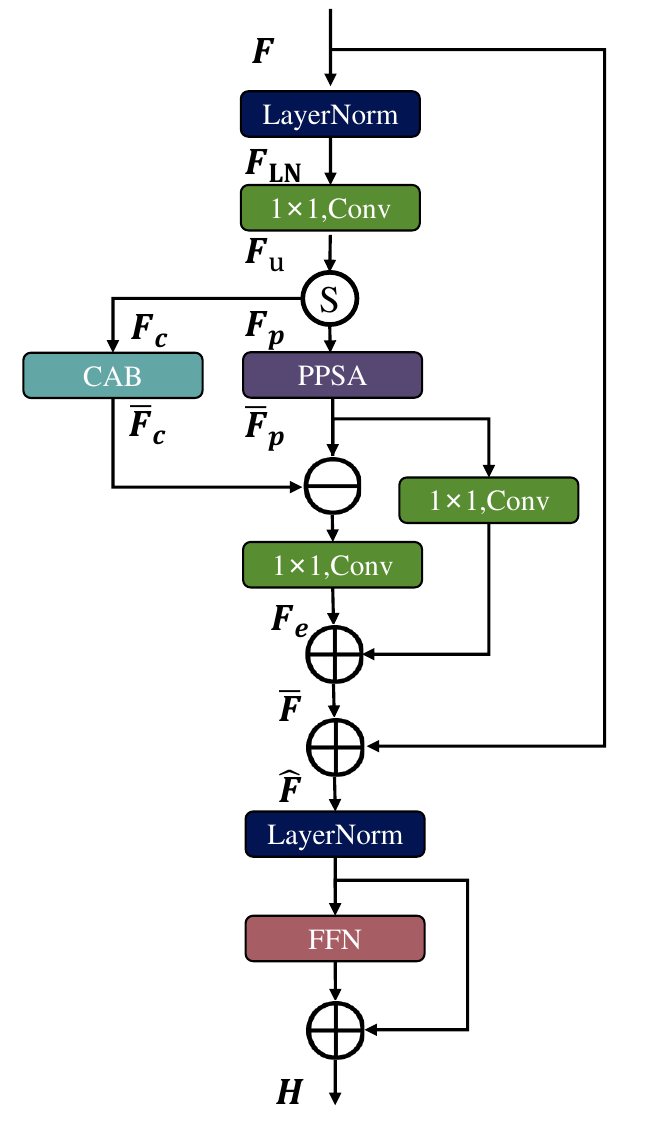}
\caption{Architectures of our proposed scale-aware backprojection-based self-attention layer.}
\label{fig_contrast}
\end{figure}

Moreover, previous studies have indicated that Transformers, which excel at modeling long-range dependencies\cite{li2023feature}, tend to focus on learning low-frequency features, contrasting with the characteristics of convolutional layers that excel at local feature modeling. To this end, building upon the powerful multi-scale feature learning capability of PPSA and the strong local modeling capability of convolutional layers, we introduce the first granularity of back-projection learning in MBT, i.e., SPAL, as shown in Fig.\ref{fig_contrast}.

SPAL consists of two parts: backprojection-enhanced PPSA and a feed-forward network. Firstly, the given feature $\bm{F} \in\mathbb{R}^{H\times W\times C}$ is preprocessed through Layer Normalization (LN) to obtain $\bm{F}_{\text{LN}}$ \footnote{For the sake of simplicity, we omitted the index number of SPAL.}. Then, a $1\times 1$ convolutional layer is applied to increase the channel dimension of the feature to $C_{1}$ in order to enhance its representational capacity, resulting in $\bm{F}_{\text{u}} = \mathcal{F}_{1\times1}(\bm{F}_{\text{LN}})$. Subsequently, the feature is split into two parts according to the channel dimensions, $\bm{F}_c\in\mathbb{R}^{H\times W\times \frac{C_{1}}{2}}$ and $\bm{F}_p\in\mathbb{R}^{H\times W\times \frac{C_{1}}{2}}$. $\bm{F}_p$ 
undergoes multi-scale learning via PPSA~$\mathcal{F}_{\text{PPSA}}$ which serves as the main branch, producing the global feature $\bm{\bar{F}}_p$ as 
\begin{equation}
\bar{\bm{F}}_{\text{p}} = \mathcal{F}_{\text{PPSA}}(\bm{F}_{\text{p}}).
\end{equation}

On the other hand, $\bm{F}_c$ is processed through a channel attention block $\mathcal{F}_{\text{CAB}}$ following HAT\cite{chen2023activating} to obtain high-frequency local features for feedback supplementation as
\begin{equation}
\bar{\bm{F}}_{\text{c}} = \mathcal{F}_{\text{CAB}}(\bm{F}_{\text{c}}).
\end{equation}
$\bm{\bar{F}}_c$ is then subtracted from $\bar{\bm{F}}_p$ and enhanced through a $1\times 1$ convolutional layer to obtain feedback error features $\bm{F}_e$ as

\begin{equation}
\bm{F}_{\text{e}} = \mathcal{F}_{{1\times1}}(\bm{\bar{F}}_{\text{c}} - \bm{\bar{F}}_{\text{p}}).
\end{equation}

By adding $\bm{F}_{e}$ to the adaptively adjusted $\bm{\bar{F}}_p$ through a $1\times 1$ convolution layer, we obtain the enhanced feature $\bm{\bar{F}}$ in a backprojection way as

\begin{equation}
\bm{\bar{F}} = \mathcal{F}_{1\times1}(\bm{\bar{F}}_p) + \bm{F}_e.
\end{equation}

Finally, a residual operation is applied by adding the input feature $\bm{F}$ to $\bar{\bm{F}}$ to alleviate training difficulties, resulting in the back-projection-enhanced feature $\bm{\hat{F}}$:
\begin{equation}
\bm{\hat{F}} = \bm{\bar{F}} + \bm{F}.
\end{equation}


The feed-forward network consists of two linear layers with an activation function between them to capture higher-level features and contextual relationships as
\begin{equation}
\bm{H} = \text{FFN}(\text{LN}(\hat{\bm{F}})) + \hat{\bm{F}}. 
\end{equation}

Unlike the Back-Projection Block, the Back-Projection structure we designed minimally introduces additional parameters and computational burden while preserving the original feedback learning. We achieve this by subtracting the rich multiscale low-frequency features learned by the PPSA branch from the high-frequency features enhanced by the channel attention block (CAB) branch and reinforcing the feature learning fed back to the PPSA branch. This combination enables the network to learn from features at different levels, enhancing feature richness and ultimately yielding superior reconstruction results.


\subsection{Context-aware Backprojection based Transformer Block}
In the SR network, cascaded residual blocks have been widely adopted for deep feature extraction~\cite{ledig2017photo,lim2017enhanced,zhang2018image}. The primary advantage of this design is that, through hierarchical feature processing, the network gains a better understanding of the image's structure and content, leading to a significant improvement in the reconstruction quality and richer detail representation. However, typically, each residual block is merely connected through a straightforward cascading mechanism, failing to effectively propagate the rich features from one residual block to the next. Such a connection scheme results in inefficient feature propagation in deep network layers, thereby impeding the achievement of more precise reconstruction results. Furthermore, due to the absence of a dedicated feature transmission mechanism, this straightforward cascading approach struggles to handle complex feature relationships, particularly when dealing with images containing intricate features, such as remote sensing images, thereby imposing certain limitations on performance. To this end, we incorporate the back-projection learning from SPAL into the context-aware backprojection-based transformer blocks (CPTBs) to enhance feature interaction between CPTBs. This constitutes the second granularity of backprojection learning proposed in MBT, as shown in Fig.\ref{fig_net}. 

Each CPTB consists of $N$ cascaded SPALs and a $3\times 3$ convolutional layer used for feature aggregation. By introducing the backprojection feature enhancement structure for each CPTB, we effectively model the feature relationships between the previous CPTB and the current CPTB, optimizing the feature propagation process between CPTBs, and allowing the network to more fully transmit and utilize deep features.

Given the feature $\bm{H}^{n-1}\in\mathbb{R}^{H\times W\times C}$ generated by the $n\!-\!1$th CPTB, a $1\times 1$ convolutional layer is first applied to reduce the computational and parameter burden within the CPTB. This $1\times 1$ convolutional layer increases the dimension of feature $\bm{H}^{n-1}$ to $C_{2}$, resulting in the initial feature $\bm{H}^{n}_{\text{init}}\in\mathbb{R}^{H\times W\times C_{2}}$. After that, a channel split operation is performed and the $\bm{H}^{n}_{\text{init}}$ is split as  $\bm{H}^{n}_{\text{p}}\in\mathbb{R}^{H\times W\times \frac{C_{2}}{2}}$ and $\bm{H}^{n}_{\text{c}}\in\mathbb{R}^{H\times W\times \frac{C_{2}}{2}}$. $\bm{H}^{n}_{\text{p}}\in\mathbb{R}^{H\times W\times \frac{C_{2}}{2}}$  is processed with $N$ cascaded SPALs and a $3\times 3$ convolutional layers for feature aggregation as 

\begin{equation}
    \bm{\bar{H}}^{n}_{\text{p}}=\mathcal{F}_{3\times3}(\mathcal{F}^{N}_{\text{SPAL}}(\mathcal{F}^{N-1}_{\text{SPAL}}(\cdots(\bm{H}^{n}_{\text{p}})\cdots))).
\end{equation}
While, for $\bm{H}^{n}_{\text{c}}$, a channel attention block is also used to enhance the feature as

\begin{equation}
    \bm{\bar{H}}^{n}_{\text{c}} = \mathcal{F}_{\text{CAB}}(\bm{{H}}^{n}_{\text{c}}),
\end{equation}
which is consistent with the design in SPAL. After that, we obtain the differential features with powerful representational capabilities as
\begin{equation}
    \bm{{H}}^{n}_{\text{e}} = \bm{\bar{H}}^{n}_{\text{p}} -\bm{\bar{H}}^{n}_{\text{c}}.
\end{equation}
The results will be further enhanced with a residual connection to obtain the complementary information as
\begin{equation}
    {\bm{\bar{H}}^{n}}=\mathcal{F}_{1\times1}(\bm{{H}}^{n}_{\text{e}})+\mathcal{F}_{1\times1}(\bm{\bar{H}}^{n}_{\text{p}})
\end{equation}
Finally, the output of the $n$-th CPTB can be obtained as
\begin{equation}
     {\bm{H}^{n}}= {\bm{\bar{H}}^{n}}+ {\bm{H}^{n-1}}.
\end{equation}

{By introducing CPTB, MBT can comprehensively capture subtle details within features and correlations between features, thus extracting deeper features that are more favorable for feature reconstruction.}

\subsection{Backprojection-based Reconstruction Module}


The mainstream super-resolution networks~\cite{lei2021hybrid, lei2019coupled, haut2019remote, qin2020remote} directly upsample the low-resolution features to obtain high-resolution images.
However, obtaining a reconstructed image solely through upsampling operations without further enhancing feature representation capability increases the training difficulty and restricts the improvement in super-resolution performance, especially when larger magnification factors are required. To this end, inspired by~\cite{liu2019image}, we introduce the third granularity of backprojection learning, referred to as backprojection-based reconstruction module, and apply it after the reconstruction layer as illustrated in Fig\ref{fig_net}. 


\begin{table*}[!t]
\begin{center}
\caption{MEAN EVALUATION METRICS OVER THE UCMERCED and AID TEST DATASET. The best and second-best results are highlighted in red and blue, respectively. The best and the second best results are marked with \textcolor{red}{red} and \textcolor{blue}{blue} fonts.}
\label{tab_all}
\renewcommand\arraystretch{2}
\begin{tabular}{ c | c| c | c | c | c | c |c | c|c|c |c}
\hline
Scale& Metric & \textbf{\makecell{SRCNN\\\cite{dong2014learning}}}& 
\textbf{\makecell{VDSR\\\cite{kim2016accurate}}}&
\textbf{\makecell{DCM\\\cite{haut2019remote}}}& 
\textbf{\makecell{LGCNet\\\cite{lei2017super}}}& \textbf{\makecell{HSENet\\\cite{lei2021hybrid}}} & \textbf{\makecell{TransENet\\\cite{lei2021transformer}}} &
\textbf{\makecell{SRDD\\\cite{maeda2022image}}} &
\textbf{\makecell{FENet\\\cite{wang2022fenet}}} &
\textbf{\makecell{Omnisr\\\cite{wang2023omni}}} &
\textbf{\makecell{MBT\\(Ours)}}\\
\hline

\multirow{2}*{\makecell{UCMerced\\ $\times $2 }} 
&PNSR&33.04	& 33.95	& 34.14	& 33.54	& \textcolor{blue}{34.32}	& 34.05	& 34.25
 & 33.95 & 34.16  & \textcolor{red}{34.65} \\
\cline{2-12}
~&SSIM& 0.9181 & 0.9281	& 0.9306 & 0.9242 & \textcolor{blue}{0.9320} & 0.9294 & 0.9319 &0.9284 & 0.9303 &\textcolor{red}{0.9352} \\
\cline{2-12}
\hline

\multirow{2}*{\makecell{UCMerced\\ $\times $3 }} 
&PNSR&29.00 & 29.78 & 29.86 & 29.36 & \textcolor{blue}{30.04} &29.90 &29.92 &
29.80 & 29.99 &\textcolor{red}{30.44} \\
\cline{2-12}
~&SSIM&0.8142 & 0.8354 & 0.8393	& 0.8247 & \textcolor{blue}{0.8433} &  0.8397 & 0.8411 & 0.8379 & 0.8403 & \textcolor{red}{0.8552}\\
\cline{2-12}
\hline

\multirow{2}*{\makecell{UCMerced\\ $\times $4 }} 
&PNSR& 26.92 & 27.56 & 27.60 & 27.18 & 27.75 & 27.78 & 27.67 & 27.59 &
\textcolor{blue}{27.80} & \textcolor{red}{28.04} \\
\cline{2-12}
~&SSIM& 0.7286 & 0.7522 & 0.7556 & 0.7394 & 0.7611 & 0.7635 & 0.7609 & 
0.7538 &  \textcolor{blue}{0.7637} & \textcolor{red}{0.7725} \\
\cline{2-12}
\hline

\multirow{2}*{\makecell{AID\\ $\times $2 }} 
&PNSR& 34.74 & 35.20 & 35.35 & 35.00 & \textcolor{blue}{35.50} & 35.40 & 35.33 & 35.33 & 
\textcolor{blue}{35.50} & \textcolor{red}{35.68} \\
\cline{2-12}
~&SSIM& 0.9299 & 0.9349 & 0.9366 & 0.9327 & \textcolor{blue}{0.9383} & 0.9372 & 0.9367 & 
0.9364 & 0.9382 & \textcolor{red}{0.9400} \\
\hline

\multirow{2}*{\makecell{AID\\ $\times $3 }} 
&PNSR& 30.63 & 31.25 & 31.36 & 30.87 & 31.49 & 31.50 & 31.38 & 31.33 &
\textcolor{blue}{31.53} & \textcolor{red}{31.71} \\
\cline{2-12}
~&SSIM& 0.8380 & 0.8526 & 0.8557 & 0.8441 & 0.8588 & 0.8588 & 0.8564 & 
0.8550 & \textcolor{blue}{0.8596} & \textcolor{red}{0.8632} \\
\hline

\multirow{2}*{\makecell{AID\\ $\times $4 }} 
&PNSR& 28.51 & 29.01 & 29.20 & 28.67 & 29.32 & \textcolor{blue}{29.44} & 29.21 & 29.16 & 
29.19 & \textcolor{red}{29.56} \\
\cline{2-12}
~&SSIM& 0.7577 &  0.7746 & 0.7826 & 0.7646 & 0.7867 & \textcolor{blue}{0.7912} & 0.7835 & 
0.7812 & 0.7829 & \textcolor{red}{0.7943} \\
\cline{2-12}
\hline 
 
\end{tabular}
\end{center}
\end{table*}

Specifically, we use the bilinear interpolation to obtain an estimated low-resolution image $\bm{\hat{I}}_{LR}$ from   $\bm{\hat{I}}_{SR}$. Next, we subtract the estimated low-resolution image $\bm{\hat{I}}_{LR}$ from the input LR image $\bm{{I}}_{LR}$ to obtain feedback information, and the results are further processed with two 1$\times$1 convolution layers $f_{1\times1}$ as: 
\begin{equation}
\hat{\bm{F}}_{LR} =f_{1\times1}(f_{1\times1}((\bm{\hat{I}}_{LR}-\bm{{I}}_{LR}))).
\end{equation}

Finally, we upsample the enhanced features $\hat{F}_{LR}$ using bilinear interpolation and combine with the estimated SR image $\bm{\hat{I}}_{SR}$ to obtain the final SR image as
\begin{equation}
    \bm{I}_{SR}=\mathcal{F}_{\text{BINR}}({\hat{\bm{F}}_{LR}}) + \bm{\hat{I}}_{SR}.
\end{equation}
Leveraging a backprojection-based reconstruction module, we further enhance the quality of SR results while avoiding excessive parameter and computational overhead.

\begin{figure*}[!t]
\centering
\includegraphics[width=0.8\linewidth]{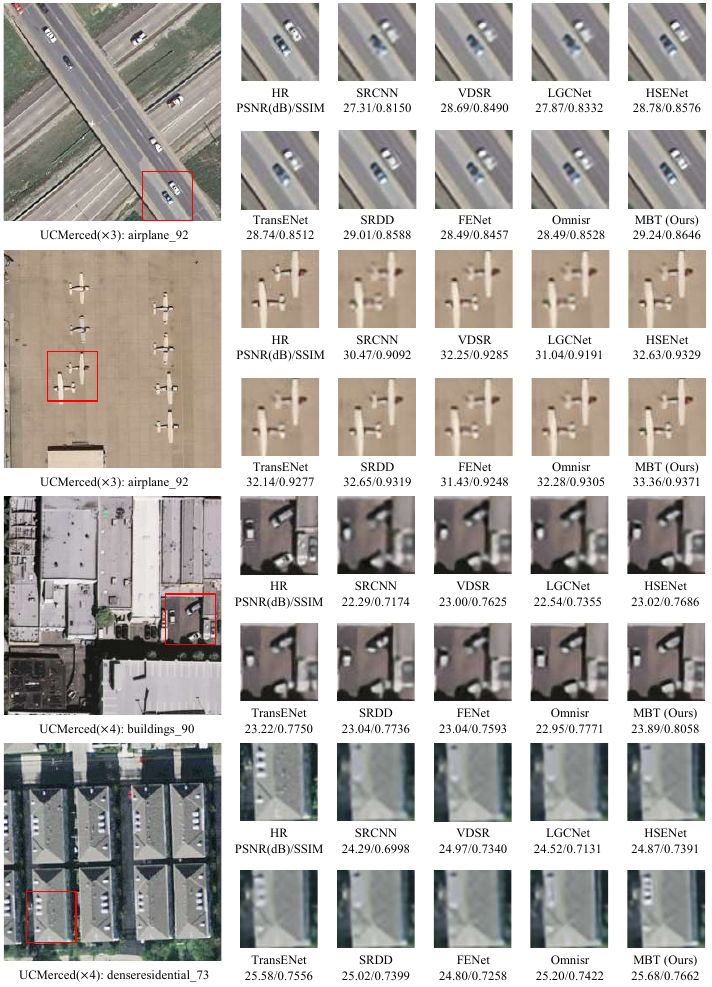}
\caption{Visual comparisons for $\times$3 and $\times$4 SR on UCMerced datasets. }
\label{fig_SR}
\end{figure*}

\begin{table}[!t]
\begin{center}
\caption{MEAN PSNR(dB) OF EACH CLASS FOR UPSCALING FACTOR 2 ON UCMERCED TEST DATASET. The best and second-best results are highlighted in red and blue, respectively.}
\label{tab_uc_x2}
\begin{tabular}{ C{12pt}|c c c c c c c}
\hline
\makebox[0.03\textwidth][c]{class} & 
\makebox[0.05\textwidth][c]{\textbf{\makecell{LGCNet\\\cite{lei2017super}}}}&
\makebox[0.05\textwidth][c]{\textbf{\makecell{TransE-\\Net\cite{lei2021transformer}}}}&
\makebox[0.05\textwidth][c]{\textbf{\makecell{Omnisr\\\cite{wang2023omni}}}}& 
\makebox[0.05\textwidth][c]{\textbf{\makecell{HSENet\\\cite{lei2021hybrid}}}}&
\textbf{\makecell{MBT\\(Ours)}}\\
\hline
1 & 32.98 &   32.90 & 32.40 &	\textcolor{blue}{32.91} & \textcolor{red}{33.99} 
\\
2 & 33.44 &	34.06 &	34.27 &	\textcolor{blue}{34.43} &	\textcolor{red}{34.79} 
\\
3 & 38.44  &	38.73 &	38.86 &	\textcolor{blue}{38.90}&	\textcolor{red}{39.04} 
\\ 
4 & 41.67  &	41.86 &	42.01&	\textcolor{blue}{42.03} &	\textcolor{red}{42.20} 
\\
5 & 32.41  &	33.19 &	33.42 &	\textcolor{blue}{33.57}&	\textcolor{red}{34.03} 
\\
6 & 29.24  &	29.41 &	29.46 &	\textcolor{blue}{29.50}&	\textcolor{red}{29.57} 
\\
7 & 32.59  &	33.29 &	33.41 &	\textcolor{blue}{33.59}&	\textcolor{red}{33.92} 
\\ 
8 & 32.84  &	32.95 &	33.05 &	\textcolor{blue}{33.06}&	\textcolor{red}{33.14} 
\\
9 & 33.62  &	34.40 &	34.57 &	\textcolor{blue}{34.78}&	\textcolor{red}{35.08} 
\\
10& 39.71  &	39.88 &	39.94 &	\textcolor{blue}{39.96}&	\textcolor{red}{40.06} 
\\
11& 29.09  &	30.30 &	30.57 &	\textcolor{blue}{30.87}&	\textcolor{red}{31.62} 
\\ 
12& 32.59  &	33.34 &	33.52 &	\textcolor{blue}{33.67}&	\textcolor{red}{34.07} 
\\
13& 32.16  &	32.71 &	32.82 &	\textcolor{blue}{32.99}&	\textcolor{red}{33.35} 
\\
14& 28.91  &	29.60 &	29.83 &	\textcolor{blue}{29.97}&	\textcolor{red}{30.41} 
\\
15& 30.66  &	31.55 &	31.64 &	\textcolor{blue}{31.92}&	\textcolor{red}{32.24} 
\\ 
16& 28.60  &	29.51 &	29.68 &	\textcolor{blue}{30.09}&	\textcolor{red}{30.61} 
\\
17& 32.84  &	33.00 &	\textcolor{blue}{33.06} &	\textcolor{blue}{33.06}&	\textcolor{red}{33.15} 
\\
18& 34.06  &   34.39 & 34.48 &	\textcolor{blue}{34.71}& \textcolor{red}{34.97} 
\\
19& 35.64  &	35.92 &	36.02 &	\textcolor{blue}{36.07}&	\textcolor{red}{36.24} 
\\ 
20& 36.45  &	37.05 &	37.24 &	\textcolor{blue}{37.33}&	\textcolor{red}{37.53} 
\\
21& 36.40  &	36.96 &	37.17 &	\textcolor{blue}{37.27}&	\textcolor{red}{37.60} 
\\ 
\hline
avg& 33.54  & 34.05 & 34.16 & \textcolor{blue}{34.32}& \textcolor{red}{34.65} 
\\
\hline 
\end{tabular}
\end{center}
\end{table}

\section{Experiment}

\subsection{Datasets}

Our experiments were conducted on two remote sensing datasets, namely, the UCMerced dataset and the AID dataset. Details of the two datasets are given as follows.

\begin{enumerate}
	\item{{\emph{UCMerced dataset\cite{xia2017aid}.}} Comprising 21 classes\footnote{All these 21 classes: 1-Agricultural, 2-Airplane, 3-Baseballdiamond, 4-Beach, 5-Buildings, 6-Chaparral, 7-Denseresidential, 8-Forest, 9-Freeway, 10-Golfcourse, 11-Harbor, 12-Intersection, 13-Mediumresidential, 14-Mobilehomepark, 15-Overpass, 16-Parkinglot, 17-River, 18-Runway, 19-Sparseresidential, 20-Storagetanks, and 21-Tenniscourt.} of remote sensing landscapes, this dataset consists of 100 images per category, each with a spatial resolution of 0.3 meters per pixel and dimensions of 256 × 256 pixels. In alignment with prior research \cite{wang2023hybrid} and \cite{lei2021hybrid}, we have evenly divided the dataset into two well-balanced subsets, each containing 1050 samples. Within the training set, 10$\%$ of the samples are reserved for validation purposes. }
	\item{{\emph{AID dataset\cite{yang2010bag}.}} Incorporating a total of 10,000 images, this dataset encompasses 30 classes\footnote{All these 30 classes: 1-Airport, 2-Bareland, 3-Baseballdiamond, 4-Beach, 5-Bridge, 6-Center, 7-Church, 8-Commercial, 9-Denseresidential, 10-Desert, 11-Farmland, 12-Forest, 13-Industrial, 14-Meadow, 15-Mediumresidential, 16-Mountain, 17-Park, 18-Parking, 19-Playground, 20-Pond, 21-Port, 22-Railwaystation, 23-Resort, 24-River, 25-School, 26-Sparseresidential, 27-Square, 28-Stadium, 29-Storagetanks, 30-Viaduct.} of remote sensing scenes. All images maintain a consistent resolution of 600 × 600 pixels, with a spatial resolution reaching up to 0.5 meters per pixel. In line with the methodology outlined in \cite{wang2023hybrid}, we partitioned the dataset into training and test sets. Specifically, 80$\%$ of the dataset was randomly selected for the training set. Within this training set, we curated validation sets by extracting five images per class. The remaining 20$\%$ of the images were set aside for use as the test set.}
\end{enumerate}

\subsection{Implementation Details}
In this paper, we explore the scale factors of $2\times$, $3\times$, and 4$\times$, and the upsampling blocks in the reconstruction part are modified according to the specific scale factor. In the training phase, the Exponential Moving Average (EMA) strategy is employed to stabilize training. $64\times 64$ image patches are randomly cropped from LR images, and their corresponding real references are extracted from HR images corresponding to the scale factor. Additionally, we augment the training images by randomly rotating them by 90°, 180°, and 270° and performing horizontal flips. We ultimately set the number of SPALs to 6, with 3 CPTBs in each SPAL. The number of channels is set to 96. The number of attention heads in the Pyramid Pooling Self-Attention layer is set to 4. The values of $C_{1}$ and $C_{2}$ are set to 96 and 64. Furthermore, Bilinear Interpolation operations are used in the backprojection-based reconstruction module for downsampling and upsampling.

We use the Adam optimizer\cite{kingma2014adam} to train our model with ${\beta _1} = 0.9$, ${\beta _2} = 0.99$, and $\varepsilon  = {10^{ - 8}}$. The initial learning rate is set to $2e^{-4}$, and the mini-batch size is 4. The total number of training epochs is 700, with the learning rate halved at 600 iterations. MBT is implemented with PyTorch\cite{paszke2019pytorch}, and all experiments are conducted on a single NVIDIA GeForce GTX 4090 graphics card.

\begin{figure*}[!t]
\centering
\includegraphics[width=0.8\linewidth]{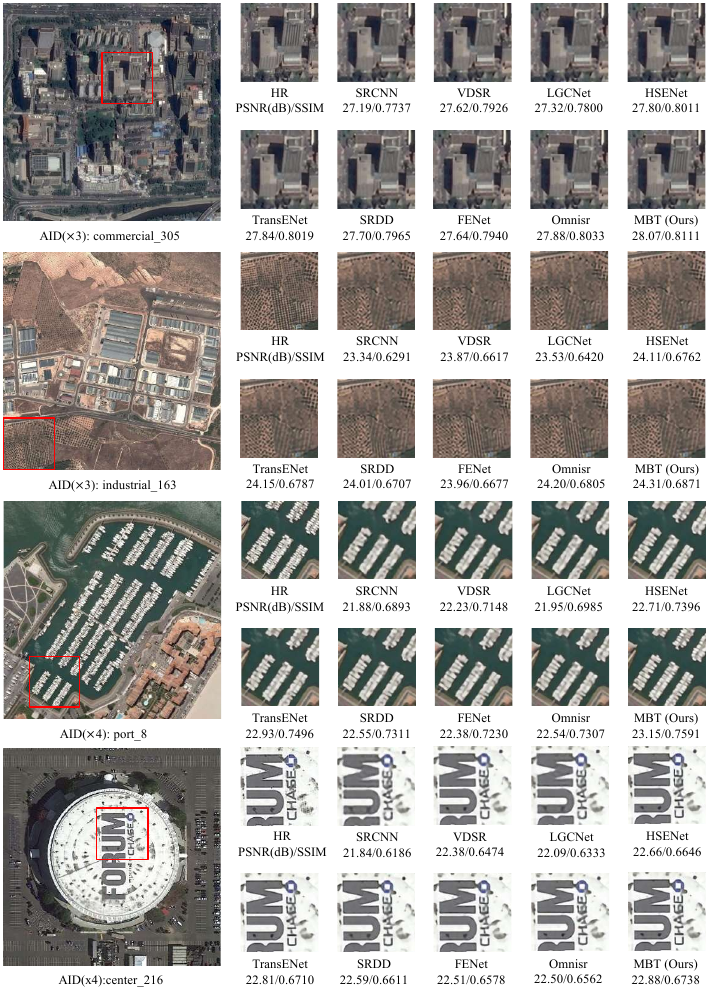}
\caption{Visual comparisons for $\times$3 and $\times$4 SR on AID datasets. }
\label{fig_SR}
\end{figure*}

\begin{table}[!t]
\centering
\caption{QUANTITATIVE COMPARISON OF PARAMETERS, MEMORY, FLOPS, AND PSNR ON UCMERCED DATASET FOR X4 SR}
\label{tab_3}
\renewcommand{\arraystretch}{1.5}
\resizebox{0.45\textwidth}{!}{%
\begin{tabular}{p{1.5cm}|c | c | c | c}
\hline
\makecell{Method}& 
\textbf{\makecell{Pub.}}&
\textbf{\makecell{\#Param}}&
\textbf{\makecell{FLOPs}}& 
\textbf{\makecell{PSNR}}\\
\hline
\makecell{LGCNet\\\cite{lei2017super}} & \makecell{IEEE GRSL\\2017}  & 0.19M  & 12.65G & 27.18
\\\hline
\makecell{DCM\\\cite{haut2019remote}} & \makecell{IEEE TGRS\\2019} & 2.17M  & 13.00G & 27.60
\\\hline
\makecell{HSENet\\\cite{lei2021hybrid}} & \makecell{IEEE TGRS\\2021} & 5.43M  & 19.20G & 27.75
\\\hline
\makecell{TransENet\\\cite{lei2021transformer}} & \makecell{IEEE TGRS\\2021} & 37.46M & 21.44G & 27.78
\\\hline
\makecell{MBT} & -- & 3.21M & 14.54G & 28.04
\\ \hline 
\end{tabular}
}
\end{table}

\subsection{Comparisons with Other Methods}
\noindent\textbf{Quantitative Results.} We compare MBT with other leading RSISR methods on UCMerced and AID datasets for $\times2$, $\times3$, and $\times4$ SR. The detailed comparisons are shown in Table I. As seen, MBT obtains the best performance in all experiment settings. Compared with CNN-based methods (e.g., SRCNN~\cite{dong2014learning}, VDSR~\cite{kim2016accurate}, DCM~\cite{haut2019remote}, LGCNet~\cite{lei2017super}, HSENet~\cite{lei2021hybrid}, SRDD~\cite{maeda2022image}, and FENet~\cite{wang2022fenet}), MBT obtains the best results. This can be attributed to the long-range dependency learning with the Transformer structure. Compared with Transformer based methods (e.g., TransENet~\cite{lei2021transformer} and Omnisr~\cite{wang2023omni})
MBT also achieves the best performance on both datasets with different up-scale factor settings, which further demonstrates the effectiveness of our proposed multi-granularity design.

Moreover, we also explore the performance of the above methods for different remote sensing classes. The comparison results are shown in Table. II. As seen, MBT outperforms other methods by significant improvements. Specifically, for the class \textit{Buildings} (\#5) and class \textit{Parkinglot} (\#16), MBT obtains even 0.46 dB and 0.52dB in PSNR compared to the second place method HSENet~\cite{lei2021hybrid}, which demonstrates the effectiveness of MBT.

\noindent\textbf{Qualitative Results.} Fig. 4 and Fig. 5 show the visual results of different methods. As seen, MBT generates the best quality visual results with clear edges and textures. This is attributed to multi-scale feature learning in SPAL and the feature enhanced using backprojection learning. The visual results demonstrate the effectiveness of MBT.

\noindent\textbf{Computational Analysis.} We also compare the network parameters and FLOPs with other RSISR methods. As shown in Table III, the parameters and FLOPs of MBT are smaller than  HSENet~\cite{lei2021hybrid}. Compared with TransENet~\cite{lei2021transformer}, MBT owns almost 8\% parameters of TransENet~\cite{lei2021transformer} but obtains the 0.26 dB PSNR improvement. 
MBT achieves a favorable trade-off between performance and model complexity.

\subsection{Ablation Studies}
\noindent\textbf{Efficacy of MBT.} We explore the main components of MBT and develop seven model variants (shown in Table IV). Our baseline model (the 1st row) removed the backprojection learning structures from CPTB and SPAL and maintained the parameter count by employing parallel calculations of CAB with adjusted channel compression ratios combined with the main pathway. Additionally, PRM was excluded. As seen, Compared to the baseline model, the individual use of SPAL (the 2nd row), CPTB (the 3rd row), and PRM (the 4th row) all achieves performance improvements. Using a combination of two modules also resulted in performance gains compared to using a single module. Particularly, the model variant that adopts SPAL and CPTB (the 6th row) achieves a performance improvement of 0.22dB in terms of PSNR compared to the baseline model, thanks to the effectiveness and rationality of SPAL and CPTB design. The full implementation model (the 8th row) obtains the best results, further demonstrating the rationality and effectiveness of the MBT design.

\begin{table}[!t]
    \centering
    \caption{\MakeUppercase{Ablation study of different model components, scale-aware backprojection based self-attention Layer (SPAL), context-aware backprojection based transformer block (CPTB), backprojection-based reconstruction module (PRM)}}
    \resizebox{0.9\linewidth}{!}{
        \begin{tabular}{c | c c c  | c}
            \hline
            \multicolumn{1}{c|}{Variant} & \textbf{SPAL}  & \textbf{CPTB} & \textbf{PRM}  & PSNR/SSIM \\ 
            \hline 
            \multicolumn{1}{c|}{1} &                &                &                    & 27.77/0.7630 \\
            \multicolumn{1}{c|}{2} & \Checkmark     &                &                     & 27.93/0.7689 \\
            \multicolumn{1}{c|}{3} &                & \Checkmark     &                     & 27.89/0.7680 \\
            \multicolumn{1}{c|}{4} &                &                & \Checkmark          & 27.86/0.7659 \\
            \multicolumn{1}{c|}{5} & \Checkmark     &                & \Checkmark          & 27.94/0.7689 \\
            \multicolumn{1}{c|}{6} & \Checkmark     & \Checkmark     &                     & 27.99/0.7716 \\
            \multicolumn{1}{c|}{7} &                & \Checkmark     & \Checkmark          & 27.90/0.7681 \\
            \multicolumn{1}{c|}{8} & \Checkmark     & \Checkmark     & \Checkmark         & 28.04/0.7725 \\
            \hline
        \end{tabular}}    
        \label{tab:ablation}%
\end{table}%

\noindent\textbf{Hyper-Parameter Settings.} We explore the performance of different numbers of CPTB in MBT, SPAL in CPTB, and the channel dimension number in MBT. The detailed results are reported in Table V. As seen, with the increases of CPTB, the performance of MBT deteriorates slightly, but the number of parameters continually increases. The reason behind the results is that too many CPTBs will increase the complexity of the model, making it difficult to achieve optimal model performance through training. Considering the trade-off between the number of parameters and the performance, we set the number of CPTB as 3.

Then, we explore the settings of SPAL in CPTB. As seen, with the increase of SPAL, the performance of MBT also improves and achieves the best results when the number is equal to 6. Thus, the number of SPAL is set to 6.

Finally, the channel dimension of MBT is also explored. As reported in Table V, MBT achieves the best results when the channel number is set to 96. Considering the trade-off between the model complexity and the final performance, we set the number of channels as 96.

\begin{table}[!t]
  \centering
  \caption{Hyper-parameter settings of MBT.}
  \resizebox{0.85\linewidth}{!}{
    \begin{tabular}{c|c|r|r|r}
    \hline
    \multicolumn{1}{c|}{Aspect} & \multicolumn{1}{c|}{\#} & \multicolumn{1}{l|}{\#Param} & \multicolumn{1}{l|}{PSNR} & \multicolumn{1}{l}{SSIM} \\
    \hline
    
          & 2     & 2.59M      & 27.97dB      & 0.7703\\
          & 3     & 3.21M      & 28.04dB      & 0.7725 \\
    \multicolumn{1}{c|}{CPTB}           
          & 4     & 3.84M      & 27.99dB      & 0.7720 \\
          & 5     & 4.46M      & 28.01dB      & 0.7720 \\
          & 6     & 5.08M      & 28.00dB      & 0.7717 \\
    \hline
    
          & 2     & 1.37M      & 27.88dB     & 0.7657  \\
          & 4     & 2.29M      & 28.02dB      & 0.7714 \\
    \multicolumn{1}{c|}{SPAL}           
          & 6     & 3.21M      & 28.04dB      & 0.7725 \\
          & 8     & 4.13M      & 28.03dB      & 0.7730\\
    \hline
    
          & 64    & 2.12M      & 27.99dB      & 0.7708 \\
    \multicolumn{1}{c|}{Channels} 
          & 96    & 3.21M      & 28.04dB      & 0.7725 \\
          & 128   & 4.65M      & 28.00dB      & 0.7715 \\
    \hline
    
    \end{tabular}}%
  \label{tab:hyper}%
\end{table}%

\section{CONCLUSION}
In this paper, we propose a Backprojection style Transformer termed MBT for RSISR. Specifically, we propose a scale-aware Transformer layer SPTL as the basic feature extraction layer for obtaining multi-scale image features. With SPTLs, we construct the Transformer block named CPTB for efficient hierarchical feature learning. Moreover, we develop a reconstruction module RPM to generate comprehensive reconstruction features for final high-resolution image reconstruction. Based on the above components, MBT is constructed and achieves promising results compared to other state-of-the-art methods on commonly-used RSISR datasets. However, other image super-resolution tasks, like arbitrary-scale super-resolution and super-resolution with complicated degradation models, are not explored in this paper, which is also important for remote sensing applications. We will extend MBT to these tasks in future work.

\bibliographystyle{IEEEtran}
\bibliography{remoteref}

\end{document}